\documentclass{article}
\usepackage{arxiv}
\usepackage{geometry}
\usepackage{hyperref}
\usepackage{cite}

\usepackage{graphicx}
\graphicspath{{figures/}{pictures/}{images/}{./}} 

\usepackage{amsmath}
\usepackage[ruled,vlined]{algorithm2e}
\usepackage{algpseudocode} 
\usepackage{array}
\usepackage{microtype}
\usepackage{caption}
\usepackage{booktabs}                  

\title{Finding Nano-{\"O}tzi: Semi-Supervised Volume Visualization for Cryo-Electron Tomography}

\usepackage{authblk}
\author[1]{Ngan~Nguyen}
\author[1]{Ciril~Bohak}
\author[2]{Dominik~Engel}
\author[3,5]{Peter Mindek}
\author[1]{Ond\v{r}ej~Strnad}
\author[1]{Peter~Wonka}
\author[4]{Sai~Li}
\author[2]{Timo~Ropinski}
\author[1]{Ivan~Viola}

\affil[1]{King Abdullah University of Science and Technology (KAUST), Saudi Arabia. E-mails: \{ngan.nguyen $\vert$ ciril.bohak $\vert$ ondrej.strnad $\vert$ peter.wonka $\vert$ ivan.viola\}@kaust.edu.sa.
	\\N. Nguyen and C. Bohak are co-first authors.}

\affil[2]{Ulm University, Ulm, Germany. E-mail: \{dominik.engel $\vert$ timo.ropinski\}@uni-ulm.de.}
\affil[3]{TU Wien, Austria. E-mails: mindek@cg.tuwien.ac.at.}
\affil[4]{Tsinghua University School of Life Sciences, Bejing, China. E-mail: sai@tsinghua.edu.cn}
\affil[5]{Nanographics GmbH}
\date{}                     
\begin{document}
\maketitle

\begin{abstract}
Cryo-Electron Tomography (cryo-ET) is a new 3D imaging technique with unprecedented potential for resolving submicron structural detail. Existing volume visualization methods, however, cannot cope with its very low signal-to-noise ratio.
In order to design more powerful transfer functions, we propose to leverage soft segmentation as an explicit component of visualization for noisy volumes.
Our technical realization is based on semi-supervised learning where we combine the advantages of two segmentation algorithms. A first weak segmentation algorithm provides good results for propagating sparse user provided labels to other voxels in the same volume. This weak segmentation algorithm is used to generate dense pseudo labels. A second powerful deep-learning based segmentation algorithm can learn from these pseudo labels to generalize the segmentation to other unseen volumes, a task that the weak segmentation algorithm fails at completely. 
The proposed volume visualization uses the deep-learning based segmentation as a component for segmentation-aware transfer function design.  
Appropriate ramp parameters can be suggested automatically through histogram analysis. Finally, our visualization uses gradient-free ambient occlusion shading to further suppress visual presence of noise, and to give structural detail desired prominence.
The cryo-ET data studied throughout our technical experiments is based on the highest-quality tilted series of intact SARS-CoV-2 virions. Our technique shows the high impact in target sciences for visual data analysis of very noisy volumes that cannot be visualized with existing techniques.


\end{abstract}


\section{Introduction}\label{sec:introduction}

\begin{figure*}
    \centering
    \includegraphics[width=\linewidth]{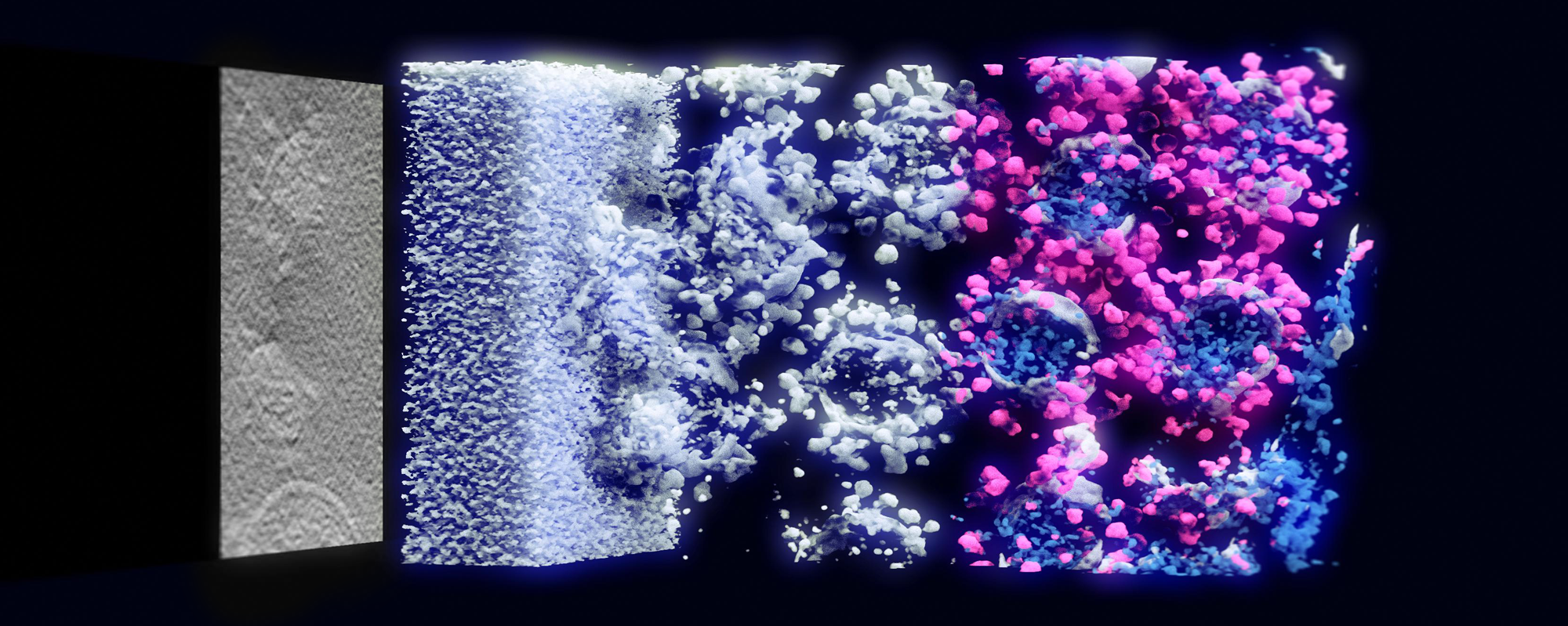}
    \captionof{figure}{From left to right: slice of the original data, direct volume rendering of the original data; foreground-background segmentation; color-coded four-class segmented data (background, spikes, membrane, lumen).}
    \label{fig:teaser}
\end{figure*}

Since 2014, the revolution in resolution~\cite{Kuhlbrandt2014} made cryogenic electron microscopy (cryo-EM) the main technique for high-resolution macromolecule structure determination. Its extension - cryo-Electron Tomography (cryo-ET) enables the use of cryo-EM for the 3D reconstruction of specimens. As it became widely accessible, the amount of acquired data has far superseded the existing data analysis pipelines' capacities. While acquisitions are still mostly done manually by microbiology experts, with the rapid increase in acquired data, new analysis tools must be developed.

One crucial analysis step is determining what specimen is represented in the acquired data and how to proceed with its processing. This is typically done using 3D data visualization methodologies. However, the imaging artifacts can make this quite challenging, and this can be seen as analogous to discovering fossils in an excavation site. Due to the fossilization, the specimens have almost the same composition as the surrounding environment, and they are difficult to distinguish. The interesting details in the cryo-ET data are similarly \emph{buried} in the surrounding noise. As the noise is a direct result of the acquisition process, where one has to be careful to evenly spread the energized particles, such that they cause as little damage to the specimen as possible, cryo-ET data always suffers from a low signal-to-noise ratio (SNR).

To \emph{excavate} the specimens from the surrounding noise, experts typically annotate the data manually and use those annotations for further steps in their research (e.g., subtomogram averaging~\cite{Leigh2019} for determining the detailed structure composition or visualization). As this process can only be undertaken by domain experts, data processing has become a bottleneck, and researchers are looking into modern automatic segmentation methods based on deep learning (DL). DL-based semantic segmentation approaches have shown great potential in a wide area of scientific disciplines, such as medicine~\cite{Litjens2017} or biology~\cite{Jumper2020}. However, when applying standard discrete DL-based semantic segmentation to cryo-ET data, the obtained crisp segmentation masks do not accurately reflect the uncertainty stemming from the imaging modality. Rather than having each pixel classified into a distinct---but possibly wrong---class, the low resolution and SNR require a higher degree of flexibility for exploring the data by the domain experts. Thus, similarly as when exploring medical volume data, it would be desirable to explore the data by means of a transfer function (TF)---a proceeding forbidden by the low SNR of cryo-ET.

Our technical contribution is centered around the key observation that a visual mapping using a transfer function specification essentially performs two tasks: 1) TF performs a \emph{soft segmentation} of objects and simultaneously 2) TF \emph{assigns optical properties}. While solving both tasks at the same time is non-trivial even for \emph{easy} noise-free modalities like medical CT data, by decomposing the role of visual mapping into two separate tasks, a solution can be found that allows a high degree of automation even for the most difficult and noise-polluted modalities such as cryo-ET.

We translate the task of soft segmentation during the visual mapping stage into the domain of probabilistic segmentation, which results in desired soft membership assignment. To achieve this, for the first time, we adopt the concept of \emph{semi-supervised learning} within the volume visualization pipeline. This methodology provides high-quality, soft segmentation even from sparse user input. With just a few supervising sparse strokes, we employ a weaker segmentation algorithm to generate dense labels, also called pseudo labels. We provide these pseudo labels to a stronger deep-learning classifier that is robust but a lot more data-hungry as it requires dense labels for training. Once trained, it provides a high-quality probabilistic (soft) segmentation for unseen datasets fully automatically.

The soft segmentation task is coupled with an optical properties assignment, which is much easier to automate when tackled separately. These two steps form together a visual mapping assignment, which results, together with advanced volume illumination models, in the excellent visual quality of rendered images. Finally, the 3D visualization can be fine-tuned by the user in case the automated methods did not find exactly the best visual parameters. Along this line, we make the following contributions:

\begin{itemize}
    \item We propose to alleviate the user from the transfer-function design task by decomposing the visual mapping task into soft segmentation and optical properties assignment. 
    \item For the soft segmentation we propose to use the concept of semi-supervised learning based on sparse user input to employ a weak classifier to obtain pseudo labels that serve as input for a strong classifier to obtaining final labels.
    \item For the transparency assignment we propose to combine the details of the raw data with the softness of the segmentation and estimate the opacity transfer function value with an iterative thresholding algorithm.
    \item We demonstrate and evaluate the concept on challenging cryo-ET data and gather expert feedback to understand the potential of the novel volume visualization pipeline.
\end{itemize}

In the following section, we place our approach into the context of the related work. We next give an overview of the proposed method and show how its individual components work together to yield a complete system and contribute to the final visualization. Sections \ref{sec:model-gen} and \ref{sec:vis-pipeline} present the novel components of our approach in detail at the reproducible level and are followed by the demonstration of our results and their evaluation. Results are discussed with two domain experts. In the final section, we conclude the paper and present possible future directions. 

\section{Related work}
Cryoelectron tomography has evolved dramatically over the course of the past few years~\cite{Koning2018} and has become a go-to method for most high-resolution \emph{in situ} structural biology challenges~\cite{Schur2019}. Extended with the subtomogram averaging~\cite{Chen2019} for obtaining even more detail for the desired structure, this is a perfect method for examining and analyzing the molecular architecture of new viruses such as the SARS CoV-2~\cite{Yao2020}.

Due to the nature of the acquisition process, the signal-to-noise ratio (SNR) is meager. Some of the reasons are: inability to create the perfect vacuum inside the acquisition tube---resulting in floating dust particles, imperfection of the acquisition sensors which do not always produce a perfect image, the inconsistencies in the preparation of the specimens, and the limitation of the energy used during the acquisition process. These and other reasons produce the noise which obscures the specimen and lowers the SNR. Huang et al. address this problem with the optimization of wavelet-based filters~\cite{Huang2018}. Shigematsu and Sigworth address this issue by analyzing different noise models~\cite{Shigematsu2013}. They all conclude that a Gaussian noise model---with preparation pipeline-dependent parameters---describes the noise in the data best.

It also proved that many existing denoising approaches based on deep learning are unsuited for cryo-ET data, as they require \emph{clean} targets without noise for training, which is currently impossible to acquire in this domain. For this reason, most existing denoising approaches are based on the Noise2Noise approach~\cite{Lehtinen2018}, which avoids this requirement by training on pairs of registered noisy data. While Noise2Noise did not consider cryo-EM data, such pairs of registered noisy cryo-EM images can be acquired using dose-fractioning, which splits the acquired electron dose in half, resulting in two independent images. Buchholz~et~al.\ use this approach to acquire such data and propose Cryo-Care~\cite{Buchholz2019}, which builds on top of Noise2Noise and provides denoising of both cryo-EM images, as well as tomograms. The later proposed Noise2Void~\cite{Krull2019} also reports promising results for cryo-EM images. Su~et~al.~\cite{Su2018} present a generative adversarial network (GAN) trained using synthetic data as \emph{clean} examples and synthetically degraded images as input to the denoiser, resulting in a model that can cope with different varieties of learned noise.

Deep learning approaches have also shown very promising results in segmentation tasks for both images~\cite{Minaee2021, Chen2020, Garcia2017, Hao2020} and volumetric data~\cite{Cicek2016, Zhu2019}. Most segmentation models are fully-convolutional neural networks consisting of an encoder and decoder part with skip connections between the encoder and the decoder. This encoder-decoder architecture is the basis for most segmentation networks, including the popular U-Net~\cite{Ronneberger2015}. The U-Net in particular is basis to many segmentation architectures that followed. While most of the architectural innovation was pioneered in 2D~\cite{Minaee2021}, many of the successful approaches can be extended to 3D. Following this recipe, the 3D U-Net~\cite{Cicek2016} was created by extending its 2D counterpart~\cite{Ronneberger2015}. Lee~et~al.~\cite{Lee2017} propose to use residual blocks in U-Nets, in addition to anisotropic convolution kernels to account for worse reconstruction quality in the z-axis. Furthermore, Etman~et~al.~\cite{Etmann2020} propose invertible U-Nets. By making the layers invertible, they can save lots of memory during backpropagation, which can be allocated to increase the network capacity, and thus performance. Siddique~et~al.~\cite{Siddique2020} provide an overview of different U-Net variants applied to different problems in the medical domain. Another relevant line of work is that of Bui~et~al.~\cite{Bui2017, Bui2018, Bui2019}, who propose different versions of a skip-connected 3D DenseNet for segmenting magnetic resonance images. Lastly, Gros~et~al.~\cite{Gros2020} propose SoftSeg. This approach is orthogonal to the above works and investigates techniques to deal with non-binary segmentation labels. Authors propose different activation functions for the output layer and adapted loss functions in order to deal with the soft labels. Using soft labels can be beneficial because they can elegantly incorporate uncertainty and inter-expert variability from the labeling process. In this work, we also make use of soft labels. Having the uncertainty in our trained network's predictions enables us to visualize the data with its uncertainty accordingly.

After discussing denoising and segmentation approaches, we also highlight some related works that use deep learning approaches to replace parts of the visualization pipeline itself. Cheng~et~al.~\cite{Cheng2019} propose to use a learned feature space for transfer function design instead of using raw intensity or first and/or second-order derivatives. Users can design transfer functions within a widget by choosing features relevant to them from an ordered feature list extracted by the neural network. DNN-VolVis~\cite{Hong2019} goes a step further and uses a neural network for the shading. Specifically, they render an unshaded image from the desired viewpoint and use an image-to-image translation network to apply shading in the style of an additionally supplied style image. Taking it a step further, Berger~et~al.~\cite{Berger2019} propose a GAN that fully synthesizes the desired renderings based on only a viewpoint and a transfer function, leaving the whole rendering process to the neural network. 

The clarity of DVR is only possible with a good definition of how the volume data translates into renderable optical properties, as defined by a transfer function (TF). The process of TF design was extensively researched in the past. The first rule-based approach to TF definition was proposed by Bergman et~al.~\cite{Bergman1995} and was used for coloring the meteorological and flow simulation volumetric data. Kindlman and Durkin~\cite{Kindlman1998} presented a semi-automatic approach for TF generation for visualizing material boundaries, taking into account intensities and their first and second derivatives. Correa and Ma~\cite{Correa2011} have later introduced a semi-automated method for generating TFs by progressively exploring TF space for maximizing visibility of important structures. Cai et~al.~\cite{Cai2013} introduce automatic TF generation using visibility distributions and projective color mapping, which matches the distribution of visible values in the current view with target one for equal pronouncement of all the features. Ljung et~al.~\cite{Ljung2016} present a thorough overview on transfer functions for direct volume rendering and present the still opened challenges. Luo and Dingliana~\cite{Luo2017} present a TF optimization based on visibility and saliency. A recent study~\cite{Ma2018} suggests the use of cell-based isosurface similarity, feature-based classification, and visibility analysis for a semi-automatic TF design.

Another very related line of work is uncertainty visualization. When visualizing predicted or approximative data, it is desirable that the user is informed about the confidence of those predictions. Prassni~et~al.~\cite{Prassni2010} visualize unconfident segmentations using sets of isolines, which naturally coincide in more certain regions, where a clear line is between foreground and background. For uncertain regions, these lines spread out, though, as the transition between foreground and background is more gradual. In the 3D visualization, they displayed uncertainty through a set of semi-transparent iso-surfaces. Lundstrom~et~al.~\cite{Lundstrom2006} propose an uncertainty-aware transfer function design. Their approach allows defining two separate 1D TFs, one for a fully certain prediction and another for a fully uncertain prediction. Depending on the actual degree of certainty of a sample classification, the two transfer functions are interpolated. Diepenbrock~et~al.~\cite{Diepenbrock2011} directly lower the saturation and value of standardized (HSV) color maps used to encode directions in fiber visualization to convey uncertainty.

\section{Technical overview}
\begin{figure}[htb]
    \centering
    \includegraphics[width=\linewidth]{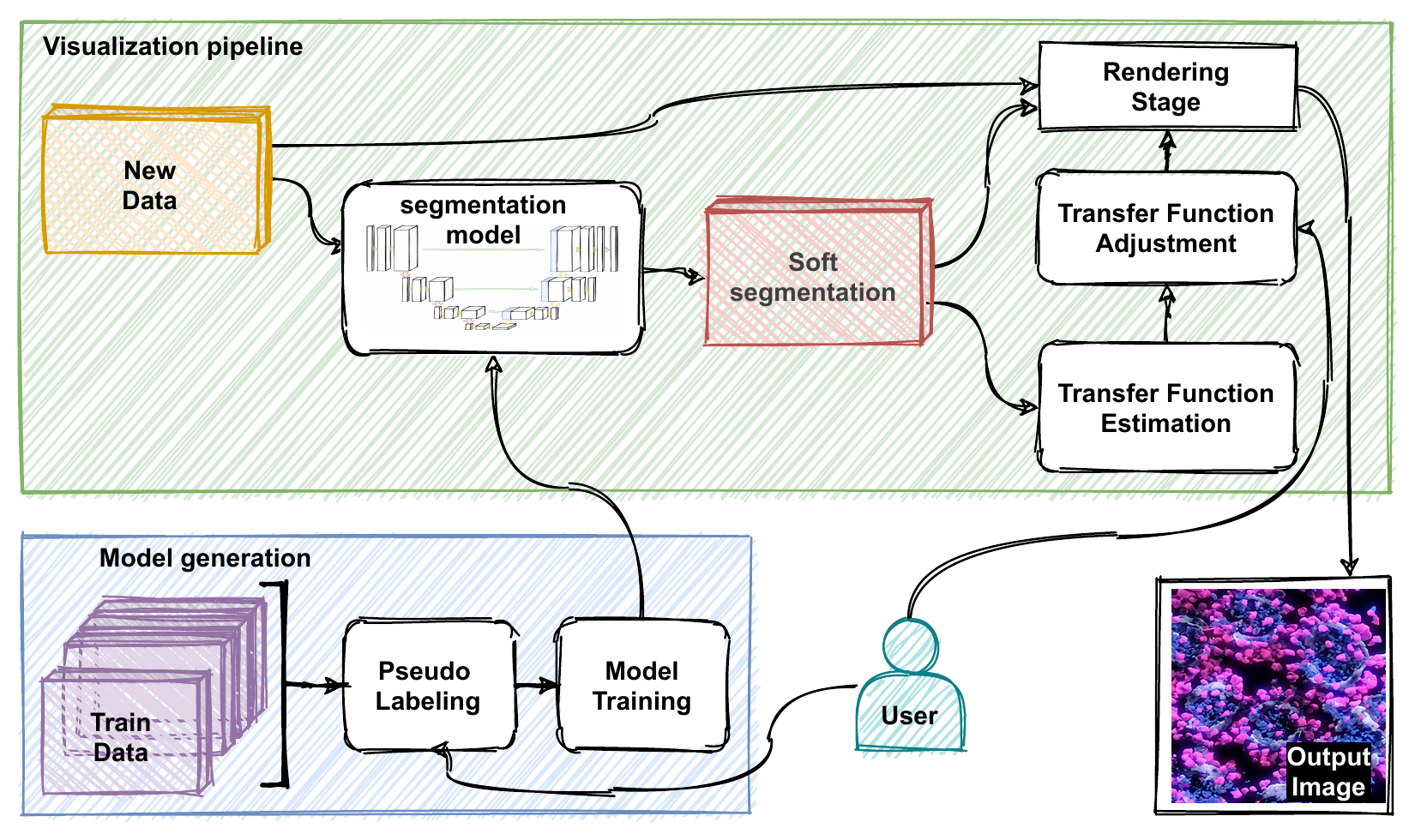}
    \caption{System overview: first sparse user-provided labels are propagated to obtain dense pseudo labels. A deep-learning based 3D segmentation model is trained on the pseudo labeled data; the resulting soft segmentation is next used in the transfer function parameter estimation step where ramping parameters are estimated; the transfer function can be further adjusted by the user before used together with the raw data and soft segmentation in the rendering stage to produce the final visualization output.}
    \label{fig:vorecem-overview}
\end{figure}

Our proposed approach enables semi-supervised direct volume rendering. We showcase that even 3D visualization of cryo-ET data becomes possible with this new approach, which is typically not achievable with current volume rendering systems, due to the low SNR (see \autoref{fig:dvr-comparison}). To achieve the desired high-quality visualization, we decompose the visual mapping stage into two sub-problems. One is automatic opacity mapping using the iterative thresholding algorithm and a mixture of soft segmentation signal with crisp raw-data signal. The second one is soft segmentation through a probabilistic approach using the semi-supervised learning methodology. Our probabilistic segmentation is composed of a deep-learning inference that performs the automatic labeling \emph{across} unseen datasets at runtime. This stage works well even on very challenging modalities, however, it is very training-data hungry. The probabilistic segmentation is therefore trained in the pre-training stage, from dense labels. 

We denote these input labels as \emph{pseudo labels} following the semi-supervised learning literature. While these labels characterize a particular volume well, the method that generates them \emph{within} the volume, performs badly in generalizing \emph{across} volumes that contain similar structures. The advantage of the pseudo-labeling method that we employ is that it produces good results for assigning soft labels \emph{within} a volume, based on only sparse user input. So one technique performs good segmentation \emph{across} volumes but requires dense input, and another method does good soft segmentation from sparse input \emph{within} a volume but does not generalize across volumes. Integrating these two approaches, the sparse-input guided segmentation with training a deep learning classifier, we obtain a semi-supervised soft segmentation concept that proves beneficial in the context of visual mapping within the volume visualization pipeline. We present our pipeline for visual mapping of two classes, i.e., foreground and background, or four classes where distinct structures of the showcased viral specimen are discriminated through distinct colors. Finally, the volume visualization incorporates integrative illumination models that further amplify the visual presence of signal over noise.

As illustrated in \autoref{fig:vorecem-overview}, the overall method consists of two parts: (1) Model generation, which takes as an input sparse expert user-annotated labels and first produces dense pseudo labels. These pseudo labels are then used by a deep-learning based segmentation algorithm for training a final 3D segmentation model. (2) Visualization pipeline, which takes new data, probabilistically segments it into four classes, estimates the transfer function parameters and renders the data.

\section{Model Generation}\label{sec:model-gen}
The final result of the \emph{model generation} step is a trained deep neural network for (probabilistic) semantic segmentation that can then be used in our proposed visualization pipeline.
In order to achieve this goal, we draw from concepts in semi-supervised learning~\cite{Lee_pseudo-label,DBLP:journals/corr/abs-1911-04252}. 
The semi-supervised learning setting applies to a situation where a smaller set of labeled data is available together with a (typically) larger set of unlabeled data. In our context, we have volumetric data. We are given a smaller set of manually labeled voxels $x_1, x_2, \ldots, x_l$ with labels $y_1, y_2, \ldots, y_l$ and a larger set of unlabeled voxels $x_{l+1}, x_{l+2}, \ldots, x_{l+u}$, where $l$ is the number of labeled and $u$ is the number of unlabeled voxels. Specifically, in our setting the labeled voxels are sparsely distributed within a given set of volumes. Since each volume is very large, it would be too time consuming to even label a single volume completely.
Our proposed solution is to use two different segmentation algorithms, leveraging the advantages from each of them. First, we use a weak segmentation algorithm provided by a state-of-the-art semi-automatic segmentation framework. The advantage of this algorithm is that it is very good in propagating segmentation information within a volume. However, it fails almost completely when propagating segmentation information across volumes. We use this segmentation algorithm to create dense pseudo labels for the remaining unlabeled voxels $x_{l+1}, x_{l+2}, \ldots, x_{l+u}$. Even though this first segmentation algorithm is quite simple, the segmentation quality within a volume is quite reasonable.
Second, we can use a more powerful deep-learning based segmentation algorithm to learn from the pseudo labels of the weak segmentation algorithm. The advantage of this algorithm is that it can learn how to generalize across different volumes. However, it is significantly more data-hungry than the weak segmentation algorithm and it is too difficult to only train it on the sparse user provided labels. The challenge in our context was to adapt existing deep learning architectures to our data and tasks. After training, the segmentation network can predict class probabilities for each of the trained classes, summing up to $1.0$, for each voxel.
Using such semi-supervised two-stage labeling approach, where manual labels are first propagated to subsets of the data---which is next used for training of a more general segmentation problem---also reflects ideas of other machine learning concepts, such as self-training~\cite{Xie2020} and distillation~\cite{Hinton2015}.
In the following subsections, we will first discuss the data to showcase the difficulty of cryo-ET data. Then we provide more details for each of the two segmentation algorithms. Finally, we also discuss data management strategies, to avoid memory issues during training and the training protocol.


\subsection{Data}
\begin{figure}[htb]
    \centering
    \includegraphics[width=0.95\linewidth]{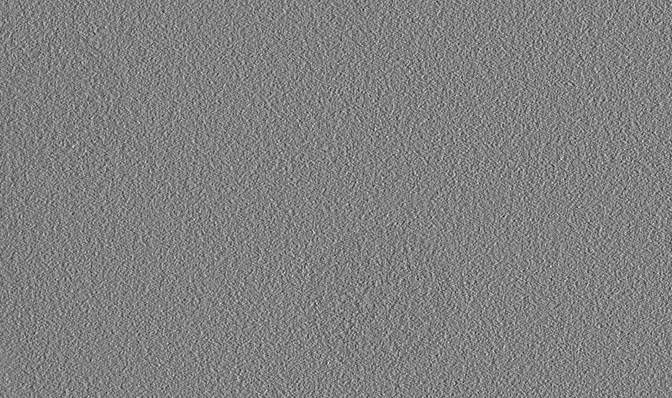}
    \caption{Single cryo-ET tomogram slice example with three virions.}
    \label{fig:data-bad}
\end{figure}

Due to the acquisition process of the cryo-ET data, where great care needs to be put into careful spreading of the electrons throughout the whole tilt series in order to avoid damaging the specimen, the acquired data is very noisy with very low signal-to-noise ratio. This is conveyed in a single slice of the volume, where individual structures can be recognized, but the objects of interest are mostly masked by the noise. An example of one slice of such data is shown in \autoref{fig:data-bad}, where three SARS-CoV-2 virions are present. While the one in the bottom middle is just perceivable to the naked eye, the other two easily blend with the surrounding noise.
Not only does this fact make images hard to segment for the untrained eye, it also makes it hard to segment. Note that directly rendering such data with common DVR techniques / transfer functions is meaningless without prior segmentation (see \autoref{fig:dvr-comparison}).  

The data used in this research consist of 60~cryo-ET volumes with a resolution of $1024\times1440\times[227-500]$ voxels. The raw data is stored with 32-bit precision resulting in 122~GB for all volumes. The pseudo label data can be stored with 16-bit precision resulting in 61.3~GB per class.
Following common best practices in deep learning, we split our 60 volumes into three sets: 50 volumes for training, 5 volumes for validation, and 5 volumes for an independent test set.

For this data, we use four classes corresponding to the \emph{spikes}, \emph{membrane}, \emph{lumen} parts, and \emph{background} of the SARS-CoV-2 virus cryo-ET data.
We also conducted experiments with two classes, \emph{background} and \emph{foreground}.


\subsection{Pseudo Label Generation}
\begin{figure}[ht]
    \centering
    \includegraphics[width=\linewidth]{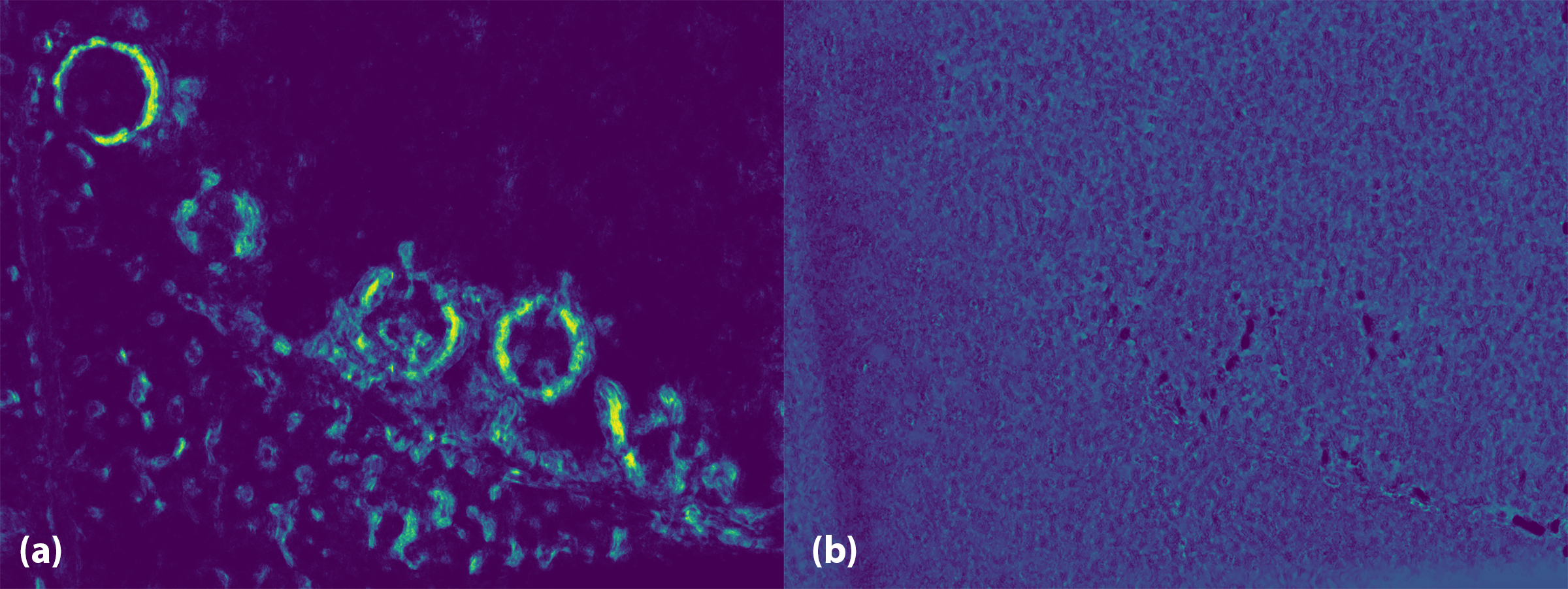}
    \caption{Comparison of a segmentation slice of the same volume using labeling parameters trained on manual annotations of the volume (a) with the use of labeling parameters estimated on another volume on this one (b) in Ilastik.}
    \label{fig:ilastik-comparison}
\end{figure}

The input to the pseudo labeling stage is the raw cryo-ET volume. The outputs are either soft or hard segmentation pseudo labels.
The pseudo labeling of the data was done using the Ilastik software~\cite{Berg2019}. We used a provided pixel classification pipeline with all 37 available 3D image features covering intensity, edge and texture properties for propagating the manual annotations throughout the volume. For each input volume we defined four classes: (1) \textit{Background}, (2) \textit{Membrane}, (3) \textit{Spikes}, and (4) \textit{Lumen}. In our foreground-background segmentation we only use the \textit{Background} class and combine the other three classes into a \textit{Foreground} class. On average, an experienced image segmentation user spends around 30 minutes for labeling a single volume. 
After the manual labeling an additional 1.5-4 hours of computation time is needed to propagate the labeled features to the whole volume and produce the soft segmentation which are also validated by the annotators. 
Ilastik works well for single-volume segmentation propagating sparse user annotations to the whole volume. However, it does not work well for propagating labeled features from one volume to other volumes, see~\autoref{fig:ilastik-comparison}. The segmentation algorithm cannot separate structure from noise. By contrast, our proposed combination of two learning algorithms drawing from semi-supervised learning produces far better results as demonstrated in the results section of this paper. The pseudo label generation stage can optionally provide soft or hard labels as output.

\subsection{Network Architecture}
To make well-informed design choices, we conducted preliminary experiments by comparing three different network architectures for foreground-background segmentation. Based on the experimental results, we selected the best-performing model.

The first architecture is the original 3D U-Net~\cite{Cicek2016}, a generalization of the 2D U-Net~\cite{Ronneberger2015} to three dimensions. It contains an encoder (contracting path) and a decoder (expanding path) part. The encoder part extracts the features from the volume, and each of its layers contains two $3\times3\times3$ convolutions each followed by a rectified linear unit (ReLU). The decoder up-samples the compressed volume back to its original resolution. Additionally, there are skip connections between the corresponding encoder and decoder layers where the resolution matches. This provides the high-resolution low-level features to the decoder.

The second architecture is the 3D Residual Symmetric U-Net~\cite{Lee2017} (3D U-Net+ResNet). It also contains the U-Net's three main components (encoder, decoder, and same-scale skip connections). To enhance the propagation of volumetric context information, each layer is set up as a residual sub-network instead of using standard convolution layers. 
In addition the authors propose to take the worse reconstruction quality along the z-axis into account, by omitting down-sampling along z, as well as by using anisotropic convolution kernels ($7\times7\times5$) to match the anisotropic nature of reconstructed volume data.

The last architecture is the Skip-connected 3D DenseNet~\cite{Bui2019} (3D DenseNet). This network also includes a contracting and expanding path. To increase the receptive field of feature maps, the contracting path contains four dense blocks. Each dense block contains four layers consisting of $3\times3\times3$ convolution, batch normalization and ReLU activation with growth rate $k=16$. There are direct connections from every layer to all subsequent layers. These connections help strengthen feature propagation. To utilize the multiple scale features in the intermediate dense blocks, the expanding path contains four 3D-upsampling operators to directly up-sample the low resolution features to the output resolution.

Experimentally, we determined that two class segmentation works better with soft pseudo labels and four class segmentation works better with hard pseudo labels.
To cope with soft labels in the foreground-background segmentation, we considered this segmentation as a regression task. Predictions represent the probability of a voxel belonging to a specific class. We experimented with different activation functions for the output layer of the network. We tested \emph{sigmoid}, \emph{softmax}, \emph{normalized ReLU}~\cite{Gros2020} and without activation, and paired them with appropriate loss functions, such as binary cross-entropy (BCE), mean squared error (MSE) and adaptive wing loss (AWL)~\cite{Wang2019}. AWL was initially proposed for heatmap regression, where regions with high intensity are usually more relevant to predict accurately, while low intensity background can be very blurry. In practice this loss tends towards using a mean squared error on the background predictions, while approaching a mean absolute error on foreground predictions. We included this loss in our experiments, because the described loss behavior is desirable in our soft segmentations as well.

\subsection{Data Management}
\begin{figure}[ht]
    \centering
    \includegraphics[width=0.8\linewidth]{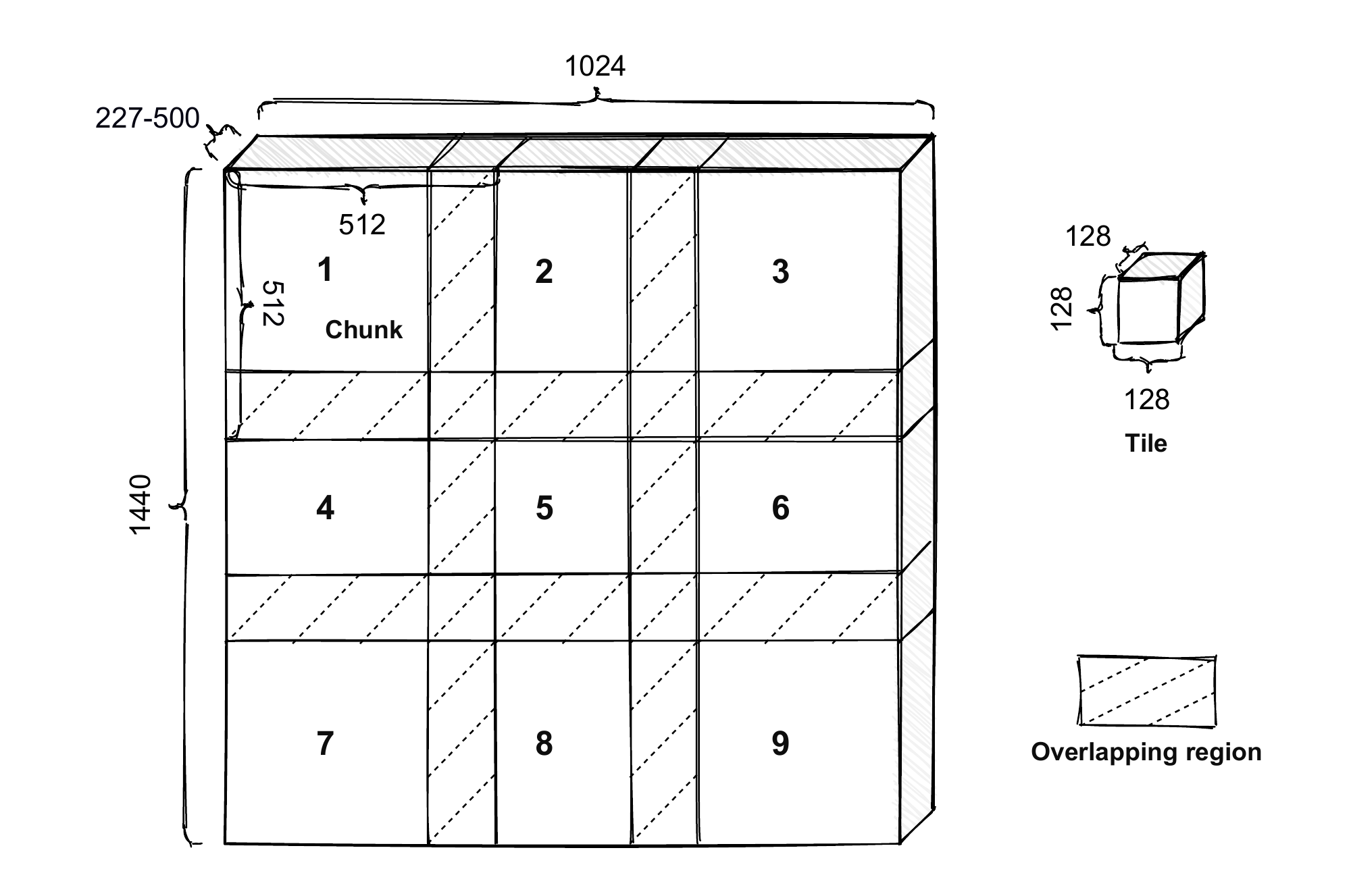}
    \caption{Scheme of dividing the individual cryo-ET volume into 9 partially-overlapping chunks.}
    \label{fig:overlapping-chunk}
\end{figure}

Due to the high resolution of our volume data, it is currently infeasible to train deep models on whole volumes directly, as we run into memory limitations. To alleviate this issue, we train our models only on tiles of size $128\times128\times128$ voxels. As we want to train on tiles in random order to prevent catastrophic forgetting, we need to load tiles from multiple volumes at once, introducing a storage bottleneck. To deal with this bottleneck, we divide each of our volumes into 9 partially-overlapping chunks of size $512\times512\times[227-500]$ prior to tiling to reduce their storage footprint, see \autoref{fig:overlapping-chunk}. Note that those chunks are significantly overlapping to avoid artifacts at the chunk borders. We also applied min-max normalization for the whole volume before splitting it into chunks. We iterate over all the chunks in each epoch during the training and randomly crop the $128\times128\times128$ input tile from each chunk. During the inference, we tile the full volume into the $128^3$-tiles with an overlap of $32\times32\times32$ voxels. To get the prediction of a whole volume from overlapping tiles, we performed alpha-blended stitching in the overlapping regions. 

\subsection{Training}\label{sec:segmentation-training}

We use PyTorch~\cite{Paszke2019} to implement the network architectures. For two class segmentation we experimented with both soft and hard labels as targets. 
For hard labels we train each network with binary cross-entropy loss. The performance of the three network architectures is reported in \autoref{table:network-comparison} and discussed in \autoref{sec:results-segmentation}. We selected the two best performing networks---3D U-Net and 3D Residual Symmetric U-Net--for the evaluation with soft labels and evaluate them using mean squared error and adaptive wing loss. The evaluation shows that the 3D Residual Symmetric U-Net with mean squared error loss performs best for such task.

Optionally, we experimented with pre-training the four class segmentation network with weights from the two class segmentation network.
In this case we can take the learned weights from the best performing two class model, omit the last layer, and transfer them to a new model used for four classes. The last layer of the new model is adapted to output four probabilities and is then fine-tuned for the four-class segmentation. This is done by changing the number of output channels of the last layer from one to four, initializing this layer using random weights, and retraining the model. We use cross-entropy loss in this step.
Note that our network still outputs soft labels (continuously-valued probabilities) that we use to visualize the segmentation certainty to the user, regardless of it being trained with hard or soft labels.
The weights of the network are optimized using the Adam optimizer $(\beta_{1} = 0.9, \beta_{2} = 0.999)$ with a batch size of~$4$. We use a learning rate of~$0.001$ and weight decay of~$0.0001$ for regularization, and mixed-precision training~\cite{Micikevicius2017} to alleviate memory limitations.

\section{Visualization Pipeline}\label{sec:vis-pipeline}
The visualization pipeline leverages the neural network trained in the model generation stage for obtaining probabilistic segmentations of new volumes needed in transfer function estimation and rendering stages described below.

\subsection{Opacity Transfer Function Estimation}
\begin{figure}[hb]
    \centering
    \includegraphics[width=0.5\linewidth]{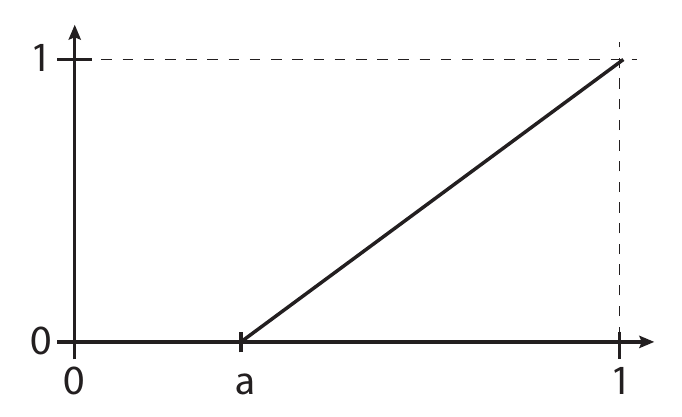}
    \caption{Transfer function ramp function. Value (a) is estimated for each class separately.}
    \label{fig:ramp}
\end{figure}

In our visualization we are combining the soft segmentation and raw input data. While we could try to extract some local geometric features from the raw data, we cannot do the same for the segmentation. The segmentation is obtained with deep neural networks from pseudo labels and we cannot rely that the local geometric features predicted by such model reflect the corresponding geometric features in the original raw data. That is why we do not want to rely on any complex transfer function design process which takes into account such features and/or maybe gradient information. We follow the simple ramping approach which proves to produce good results while not relying on the complex properties. In addition to that, the simplicity of the ramp transfer functions allows  simple modification and fine-tuning by the domain experts after an initial transfer function configuration is estimated automatically. After manual experimentation we realize that the right limit of the ramp function should be at the end of the fuzzy value interval of $[0.0,1.0]$, at $1.0$ to obtain the best visual results, leaving only the left limit of the ramp as an unknown parameter that we will estimate. To find this ramp parameter we use a simple iterative image thresholding technique~\cite{Ridler1978} outlined in Algorithm~\ref{alg:ridler}.
\begin{algorithm}
\SetKwRepeat{Do}{do}{while}
\SetAlgoLined
	\caption{Automatic Thresholding} 
	\SetKwInOut{Input}{Input}
    \SetKwInOut{Output}{Output}
    \SetKwProg{AutomaticThresholding}{AutomaticThresholding}{}{}
    
    \AutomaticThresholding{$(image)$}{
        \Input{Image - $image$}
        \Output{Threshold - $threshold$} 
        \texttt{\\}
        $T = []$, $i = 0$\;
        $imHist = histogram(image)$\;
        $meanInt = Mean(imHist)$\;
        $T[i] = Round(meanInt)$\;
        \Do{$(Abs(T[i] - T[i - 1]) \geq 1)$}{ 
            $meanIntBelowT = imHist < T[i]$\;
            $bgIntegrator = Mean(meanIntBelowT)$\;
            $meanIntAboveT = imHist \geq T[i]$\;
            $fgIntegrator = Mean(meanIntAboveT)$\;
            $i = i + 1$\;
            $T[i] = Round((bgIntegrator + fgIntegrator)/2)$\;
        }
        $threshold = Normalize(T[i])$\;
    }
\label{alg:ridler}
\end{algorithm}
The thresholding algorithm is applied to all slices of a segmented volume individually and the mean of the threshold values is taken as the left limit (a) of the ramp function $r_i$ (see \autoref{fig:ramp}) for $i$-th class:
\begin{equation}
    r_i(s) = \min\left(\max\left(\frac{p_i(s) - a}{1.0 - a}, 0.0\right), 1.0\right),
\end{equation}
where $s$ is sample along the ray, $p_i(s)$ is the class probability for the $i$-th class at the sample location $s$ in the corresponding volume, $a$ is the left limit of the ramping function.

It shows that slice-based automatic thresholding does not return an equal value for all the slices of the volume. We found that the thresholds decrease for slices towards the top and bottom of the volume. We have investigated the option of addressing this phenomena in our transfer function---as the threshold changes can be nicely approximated with a simple quadratic function---but there is no significant improvement in the final visualization, making the use of a more complex transfer function less prominent.

\begin{figure}[h]
    \centering
    \includegraphics[width=\linewidth]{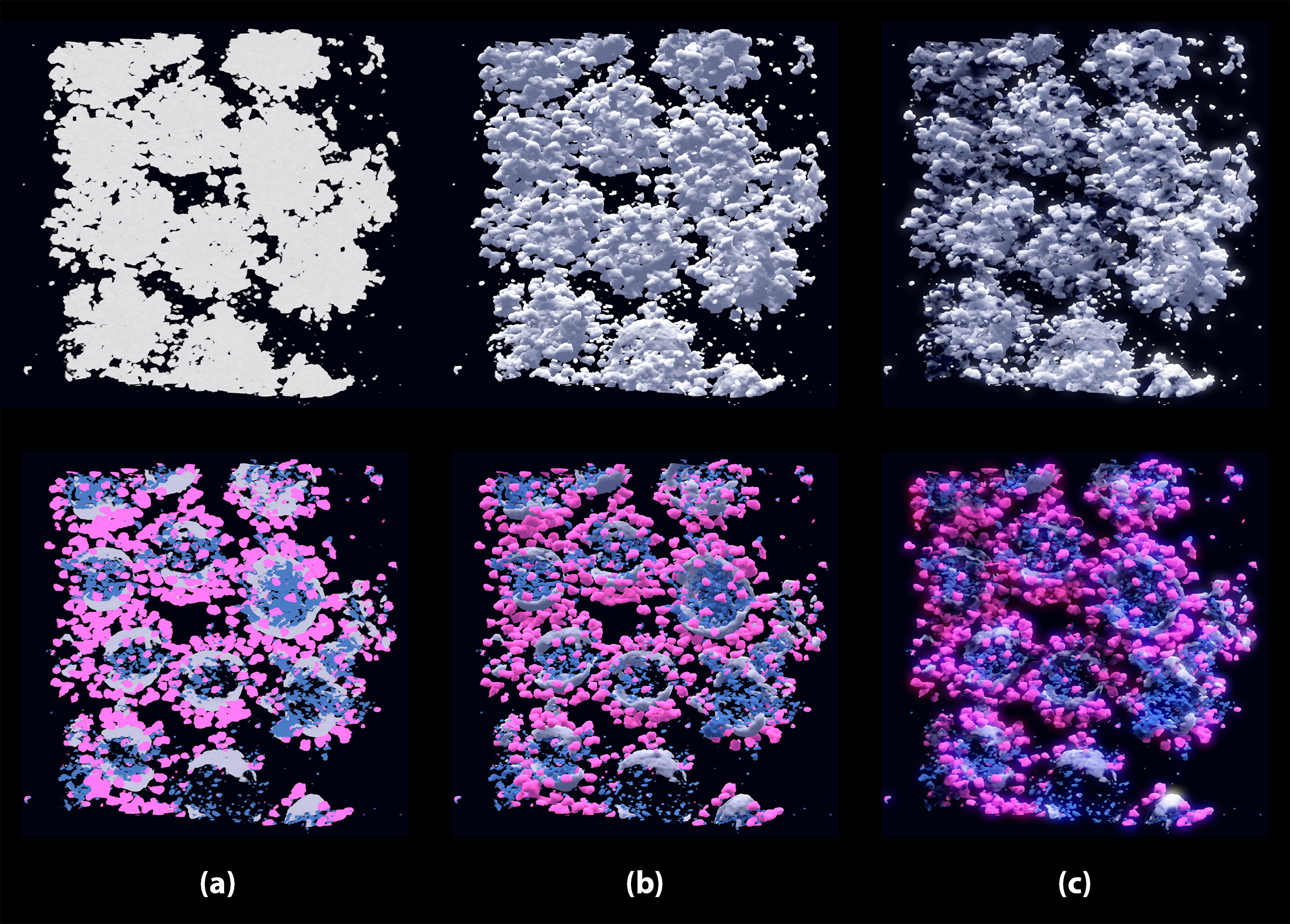}
    \caption{The figure shows how the foreground-background data (top row) and four-class segmented data (bottom row) is rendered in different stages of the pipeline: (a) material color only, (b) added local ambient occlusion, and (c) added soft shadows and bloom. In all visualization the respective masks are multiplied by the low-pass filtered original data.}
    \label{fig:rendering-pipeline}
\end{figure}

\subsection{Rendering}
Our goal is to achieve real-time speeds for the whole volume rendering, which limits the selection of the volumetric rendering technique. As an input we take raw cryo-ET volume $V_{raw}$, three soft segmentation volumes $V_{i}$, user-defined segmentation class colors $c_i$, and estimated OTF parameters presented above defined with corresponding function $r_i$. The raw volume $V_i$ is inverted and low-pass filtered to pronounce the structures in the preprocessing step, since the structures in the original data are represented with lower values. 

In order to produce meaningful and clear visualizations of the fuzzy data, we avoid using normals for illumination. As normals would have to be estimated by calculating gradients, not only we would amplify the noise by using the normals, we would also increase the amount of texture fetches that would have to be performed per sample. Instead, we approximate light scattering effects by sampling areas surrounding the illuminated voxels. We use a single spherical-light for illuminating the scene, for local shadow estimation. The rendering consists of 4 stages:
\begin{enumerate}
    \setlength\itemsep{0em}
    \item \emph{Material color}: A unique user-defined base color is used for each segmentation mask, significantly different from other classes to distinguish each class in visualization---see \autoref{fig:rendering-pipeline} (a).
    \item \emph{Local ambient  occlusion}: Local ambient occlusion is calculated in object space by sampling a sphere around each voxel and calculating the sum of the sampled voxel values form all the masks~\cite{Hernell2010}. It adds local shadows that enhance the depth perception during interaction. The sampling sphere is offset upwards to simulate illumination from above---see \autoref{fig:rendering-pipeline} (b).
    \item \emph{Soft shadows}: Soft shadows are using Monte Carlo integration of directional in-scattering calculated in object space by sampling a cone starting at the given voxel, oriented towards a spherical light source~\cite{Veach1997}. The contributions of the voxels are modulated by their distance to the original voxel. The soft shadows help distinguishing individual objects from each other and further enhance the depth perception---see \autoref{fig:rendering-pipeline} (c).
    \item \emph{Postprocessing}: In the postprocessing step, we add bloom by separating the brightest parts of the image by a tone-mapping curve, blurring them with a Gaussian kernel, and adding the result to the original visualization. The bloom highlights the brightest parts of the image, increasing the overall contrast. The post-processing is executed in screen space.---see \autoref{fig:rendering-pipeline} (c).
\end{enumerate}

The first three steps of the rendering can be described by the following equations:
\begin{align*}
    c_{o} &= \sum_{s=s_0}^{s_{max}} \sum_{i=i_0}^{i_{max}} r_i(V_{i, s}) \cdot r_{raw}(V_{raw, s}) \cdot c_i \cdot \mathrm{ssc}_s \cdot \mathrm{lao}_s \cdot a_{s, i} \\
    a_{s, i} &= \prod_{k=s_0}^{s} \prod_{j=i_0}^{i-1} (1 - r_j(V_{j, k}) \cdot r_{raw}(V_{raw, k}))
\end{align*}
where $c_o$ is output color, $s_0, \dots, s_{max}$ is the ordered set of samples along the ray, $i_0, \dots, i_{max}$ are the segmentation classes, $r_i(x)$ is the ramping function for the $i$-th class, $V_{i,s}$ is $s$-th sample along the ray inside the $i$-th class volume, $c_i$ is user-defined material color for $i$-th class, $\mathrm{ssc}_s$ is soft shadow contribution at sample $s$, and $\mathrm{lao}_s$ is local ambient occlusion contribution at $s$-th sample. The second equation $a_{s,i}$ is the accumulated alpha on the ray up to sample $s$ for class $i$.

The starting values of all the left ramp limits (for all volumes) are estimated with the approach presented in the previous section, but are still user-adjustable.

Compositing of the contributions along rays cast through every pixel of the rendering canvas produces the real-time visualization demonstrated in the following section.

\section{Results}

We have tested our technique on a challenging, but high-quality imaging dataset depicting the SARS-CoV-2 virions. In total we were provided 300 cryo-ET volumes of approximately 0.5~TB in size. Due to limitations of how many volumes can be used for training and the availability of training segmentations, we have limited our experimental data set to 60 volumes. We have performed a series of experiments on how to visualize the noisy cryo-ET data and on how to obtain fuzzy interpretations that are needed for the purpose of visual mapping. We briefly summarize the outcome of these experiments below. For the segmentation results we begin by introducing the evaluation metrics used in our experiments, before outlining our architecture selection and lastly presenting the results of our best models for both soft foreground-background segmentation and segmentation into spikes, membranes, lumen and background. For the evaluation of our visualization, we compare our technique with standard visualization techniques conveying the current technological offer and the substantial improvement in visual quality.

\subsection{Segmentation Results}\label{sec:results-segmentation}
For evaluation, we use the F1 score or Dice similarity coefficient---a common evaluation metric for image segmentation---suitable for datasets that contain imbalanced class distribution. We define $TP$ as the number of true-positive predictions, $FP$ as the number of false-positive predictions, $FN$ as the number of false-negative predictions, and $TN$ as the number of true-negative predictions. The F1 score measures the similarity between labels and predictions and is defined as:
    \begin{equation}
        F1 = \frac{2TP}{2TP + FP + FN}
    \end{equation}
A higher F1 score indicates a better result. Note that the F1 score requires the use of discrete labels. To calculate an F1 score for our continuous values in the soft segmentation, we use a threshold of $0.5$. This threshold is applied for both the label and prediction.
In the 4-class case we use \emph{argmax} to discretize the predictions.

We first evaluate different network architectures using binary labels, as detailed in \autoref{sec:segmentation-training}. We compare the standard 3D~U-Net~\cite{Cicek2016} with the Residual Symmetric U-Net~\cite{Lee2017} (3D~U-Net+ResNet) and the 3D DenseNet~\cite{Bui2019} on predicting standard binary foreground - background labels on our data. \autoref{table:network-comparison} shows the results of this experiment. We show that both the standard 3D~U-Net and the 3D~U-Net+ResNet achieve similar performance, clearly outperforming the 3D~DenseNet both in terms of F1 score and training time. Based on these results we decided to further investigate the two U-Net based models.

\begin{table}[h]
\centering
\begin{tabular}{lccc}
\toprule
 \textsc{Method} & \textsc{BCE Loss}  & \textsc{F1 Score} & \textsc{Train Time}  \\
\midrule
 3D U-Net & 0.4280 & 0.6896 & 9h 10m \\ 
 3D U-Net+ResNet & 0.4360 & 0.6776 & 11h 58m \\
 3D DenseNet & 0.4701 & 0.6254 & 21h 21m \\
 \bottomrule
 
\end{tabular}
\caption{Performance on the validation set of three network architectures for binary segmentation}
\label{table:network-comparison}
\end{table}

In the next experiment, we trained the two U-Net based models for soft segmentation. In contrast to the binary label experiment, we now face a regression problem and have therefore investigated several activation functions for the output layer, and several loss functions. The results from this experiment can be seen in \autoref{table:soft-segmentation}. The comparison shows that 3D~U-Net+ResNet with MSE loss works best. The reason for this is the decoder of this network, which outperforms the regular U-Net. Also note that the F1 score is comparatively low in this experiment. This is due to the fact that we need to binarize both the soft labels and predictions in order to compute the F1 score, which may punish numerically accurate predictions that are closely around the binarization threshold.

\begin{table}[ht]
\centering
\begin{tabular}{lcccc}
\toprule
    
\multicolumn{1}{r}{\textsc{Model}} & 
\multicolumn{2}{c}{\textbf{U-Net}} &
\multicolumn{2}{c}{\textbf{U-Net+ResNet}} \\
\cmidrule(lr){2-3}
\cmidrule(lr){4-5}

\multicolumn{1}{l}{\textsc{Loss+Act}} & 
\multicolumn{1}{c}{\textsc{Loss}} &
\multicolumn{1}{c}{\textsc{F1 Score}} &
\multicolumn{1}{c}{\textsc{Loss}} &
\multicolumn{1}{c}{\textsc{F1 Score}} \\
\midrule
\textsc{MSE+None} &
0.01798 & 0.5918 & 0.01678 & \textbf{0.6064} \\ 
\textsc{AWL+NReLU} &
0.03246 & 0.4058 & 0.03663 & 0.4874 \\ 
\bottomrule
\end{tabular}
\caption{Soft segmentation results comparing different activations (\textsc{Act}) and loss functions.}
\label{table:soft-segmentation}
\end{table}

Lastly we fine-tuned our best models from the previous experiment for use with all four classes. As our four-class labels are rather decisive we fine-tune the model using discrete labels instead of soft labels.
We aimed to train a general foreground-background segmentation model for soft labels that can be used for other datasets in the same domain. In the case of SARS-CoV-2, there are 4 classes, so we fine-tuned the model for four-class labels. Another reason is that training directly on 4 classes would require too much time and resources. For foreground-background segmentation, we needed 512 GB of RAM, and with 4 GPUs it took approximately 2 hours and 15 minutes per training epoch. In the case of four-class segmentation, it would take more than 9 hours per epoch, and the batch size would have to be decreased drastically, further prolonging the training. The F1 score of this fine-tuned model on the validation set is \textbf{0.9618} and the cross entropy loss is $0.1158$.
One of the reasons why the F1 score in this case is higher than in foreground-background case because of the nature of F1 score. The F1 score is a good measure for incorrectly classified cases. In the foreground-background segmentation, there are two classes, so the F1 score penalizes a wrong prediction to a higher degree compared to the four-class segmentation.

The training of the final foreground-background model took 5 days and 16 hours to converge. The fine-tuning of the selected transfer-learned model took additional 2 days and 15 hours. The model inference takes 20-25 seconds per volume on a single Nvidia V100 GPU computing node or 10-15 minutes per volume on a workstation with single Nvidia Quadro RTX 8000 GPU. 

The generation of pseudo labels using Ilastik was done on a workstation computer with 2$\times$Intel Xeon Gold 6230R @ 2.1 GHz, 256 GB of RAM, and Nvidia Quadro RTX 8000 48 GB GPU running Microsoft Windows 10. The deep-learning experiments were performed on diverse hardware: the model selection experiments were mostly performed on IBEX---heterogeneous group of computing nodes---at KAUST, the final model optimization was performed on a single computing node with 2$\times$Intel Xeon Gold 6242 @ 2.8 GHz, 512 GB RAM, 4$\times$Nvidia Quadro RTX 8000 48 GB GPUs running Ubuntu Linux. The neural net's  inference time was measured on both, machine used for labeling, as well as on the computing cluster.

\subsection{Visualization Results}
\begin{figure}
    \centering
    \includegraphics[width=0.7\linewidth]{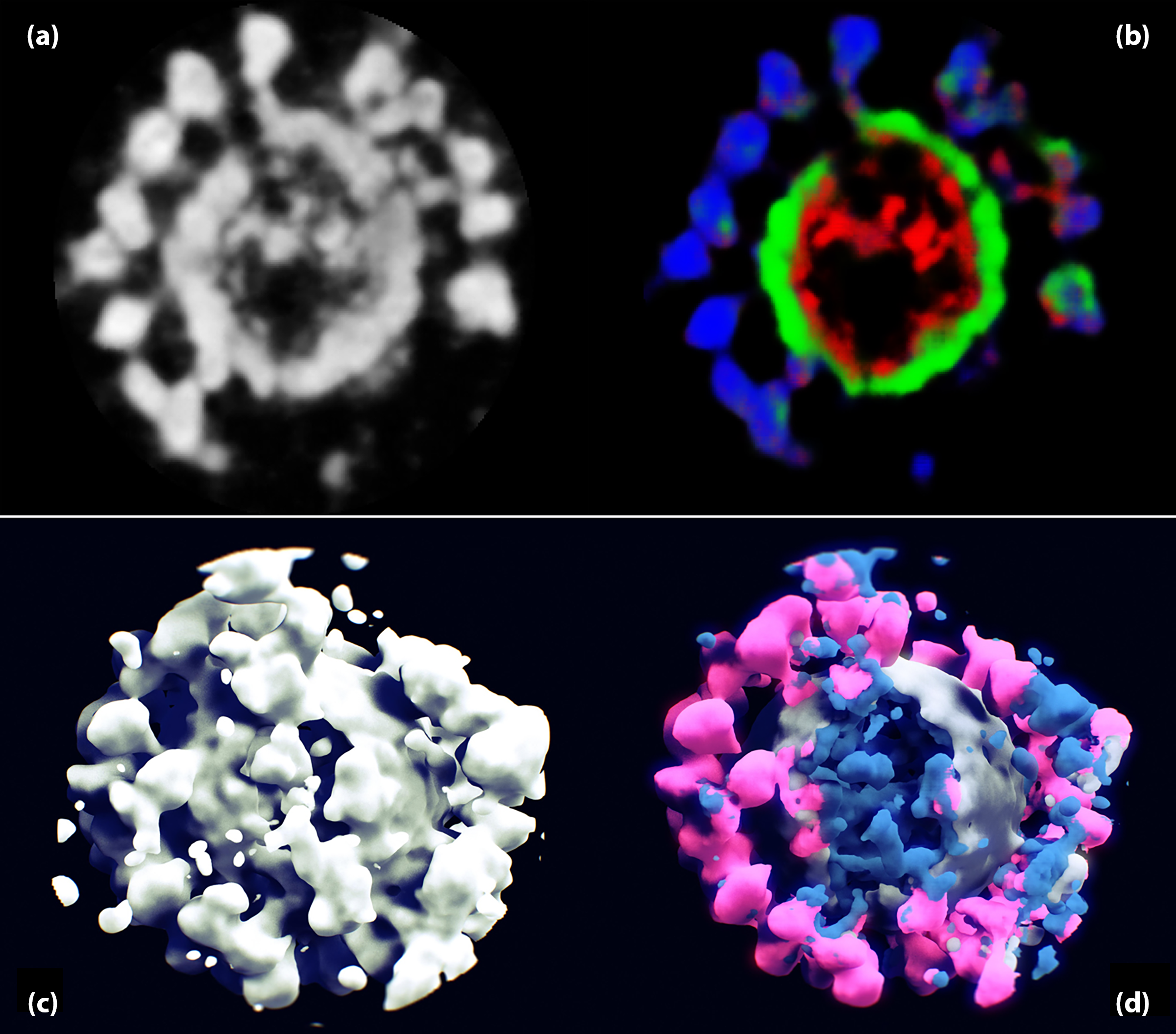}
    \caption{Figure shows one slice comparison of single virion segmented with foreground-background approach (a) and with four-class approach (b)---spikes in blue, membrane in green and lumen parts in red color. The 3D visualization of these segmentations with our rendering pipeline are shown in (c) and (d) respectively---spikes in pink, membrane in gray and lumen parts in blue color.}
    \label{fig:results-f-b-vs-four-class}
\end{figure}

The final visualizations of the proposed approach are displayed in various figures. The teaser image \autoref{fig:teaser} shows how we get from the \emph{solid} cryo-ET volume, over the foreground-background, to the four-class segmented visualization. We show segmentations and final renderings of a single virion segmented with foreground-background, as well as four-class segmentation, in \autoref{fig:results-f-b-vs-four-class}. The dimension of this single virion sub-volume is $246 \times 264 \times 340$ voxels.

For visualization purposes the input data was down-sampled to 8-bit precision. In the case of foreground-background segmentation visualization, two volumes are loaded to the GPU, consuming 0.69 - 1.37~GB of GPU memory. In the case of four class segmentation visualization, four volumes were loaded to the GPU resulting in 1.37 - 2.75~GB of memory use. We also tested our system on the full 16-bit precision---resulting in 2.75 - 5.49~GB of memory consumption. Apart from the loading times there were no other differences in the performance, i.e. the frames-per-second (FPS) count is the same. In \autoref{table:visualization-evaluation} one can see how the FPS changes with different segmentations at different resolutions.

\begin{table}[!t]
    \centering
    \begin{tabular}{lcc}
    \toprule
     \textsc{Resolution} & \textsc{Foreground-Background}  & \textsc{Four-Class}\\
    \midrule
     Full-HD & 26.24  & 21.50 \\     
     4K      & 10.89  &  9.20 \\     
     \midrule
     \textsc{VRAM Cost} & 1.37 - 2.75 GB & 2.75 - 5.49 GB \\
     \bottomrule
     
    \end{tabular}
    \caption{Rendering performance evaluation for both segmentations on two resolutions, given in frames per second. Additionally we report the range of GPU memory usage from the smallest to largest volumes.}
    \label{table:visualization-evaluation}
\end{table}

We ran an automated visualization task of 360° tilt rotation of all five test volumes for foreground-background segmentation (2 volumes), and for four-class segmentation (4-volumes). Each experiment was run for five times on Full-HD (1080p) and 4k (2160p) resolution, respectively, then the measurements were averaged. 
The visualization evaluation was done on a workstation computer with 2$\times$Intel Xeon Gold 6230R @ 2.1 GHz, 256 GB of RAM, and Nvidia Quadro RTX 8000 48 GB GPU running Microsoft Windows 10.

\begin{figure}[htb]
    \centering
    \includegraphics[width=\linewidth]{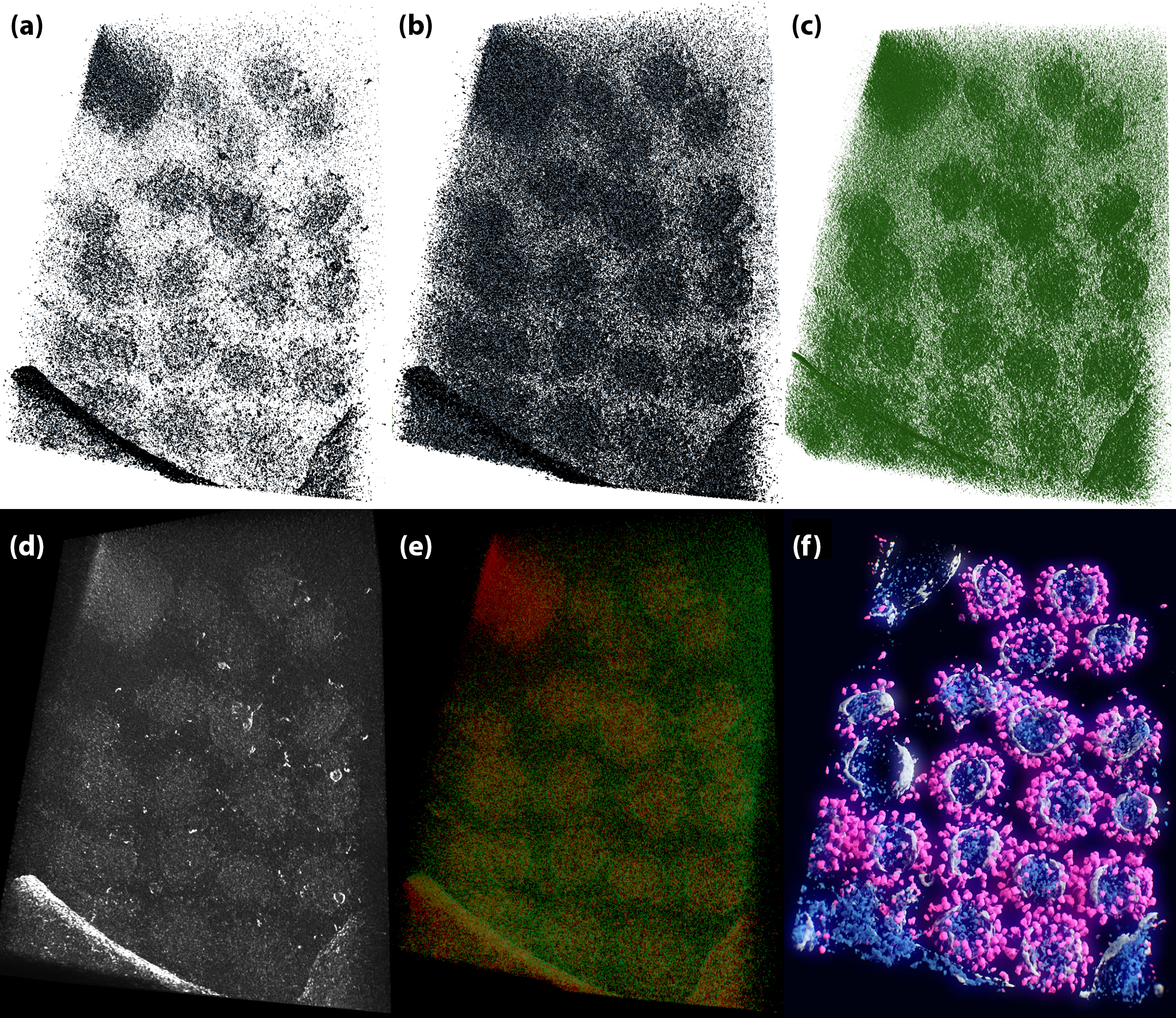}
    \caption{Comparison with regular DVR rendering techniques: (a) and (b) ISO surface rendering with 2 different ISO values, (b) regular DVR, (c) VPT, (d) MIP, (e) color TF DVR, and (f) our approach.}
    \label{fig:dvr-comparison}
\end{figure}

\begin{figure}
    \centering
    \includegraphics[width=\linewidth]{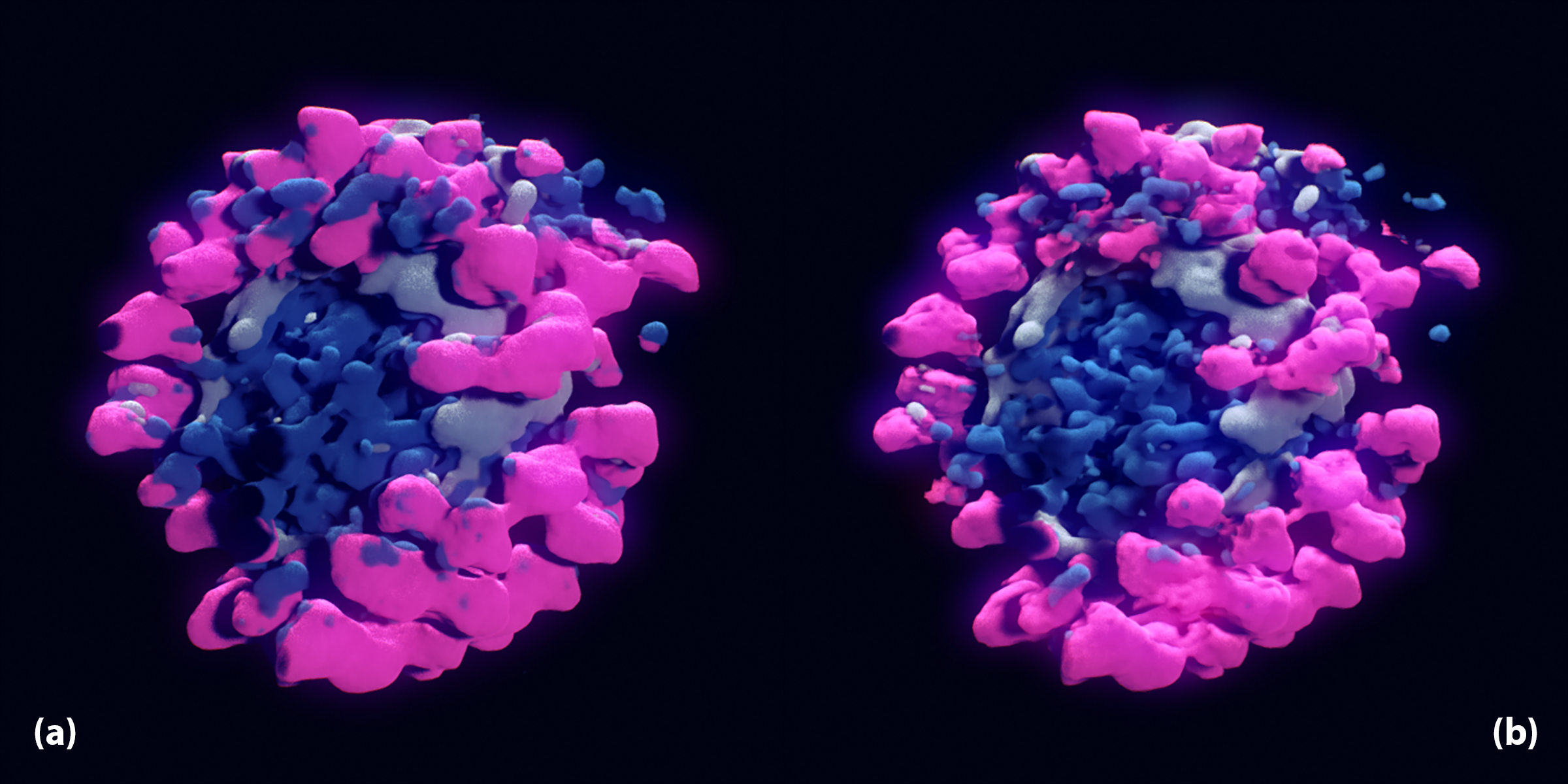}
    \caption{Comparison of our method without multiplication of segmentation volumes with original data (a) and with multiplication (b).}
    \label{fig:dvr-multiplication}
\end{figure}

To show how hard it is to render the cryo-ET data directly we have used regular volume rendering approaches: ISO surface rendering, direct volume rendering (DVR), maximum intensity projection (MIP), and volumetric path tracing (VPT). All these methods were used for rendering on a 2k canvas and run in real-time, except the VPT, where one needs to wait for the convergence. Since the results on the original data were unusable, we show the renderings performed on the low-pass filtered data that we also use in our approach to suppress high-frequency noise and pronounce the specimen structures. The visual comparison with other DVR techniques is displayed in \autoref{fig:dvr-comparison}. While one can see basic shape outlines---even with this with low-pass filtered data---the details are unrecognizable. There is also no easy way to configure the 2D transfer function to distinguish between the four classes we segment for in our approach.
In \autoref{fig:dvr-multiplication} we show the impact of the inclusion of the original data in the visualization. Specifically, we multiply the original data with the segmentation volumes to reveal the fine details and structures of the original data.

\section{Discussion}
We have discussed the results of our work with two domain experts. One domain expert is the co-author of this paper, and the other one is an independent domain expert.

The first expert is a physicist specialized in biophysics. He has 12 years of experience in the field and 8 years in cryo-ET. He is head of the cryo-ET laboratory at his University and works with cryo-ET data almost every day. He works with biological specimens only and in this paper he is among the co-authors.

He is in charge of data acquisition used in this study. He confirms that the semi-automatic segmentation (SAS)---our pseudo labels--- is good but could be improved, especially for the lumen structure annotations, where specialized domain knowledge would be beneficial. Moreover, he suggested that semi-automatic segmentation might be easier on preprocessed data, which should be investigated in the future. He confirms that the results are satisfactory for the amount of time spent on the segmentation task and that full manual annotation would take up to several days per volume. He suggests considering omitting the top-most and bottom-most parts of the volumes where there are artifacts due to an air-water boundary.

\begin{figure}[h]
    \centering
    \includegraphics[width=\linewidth]{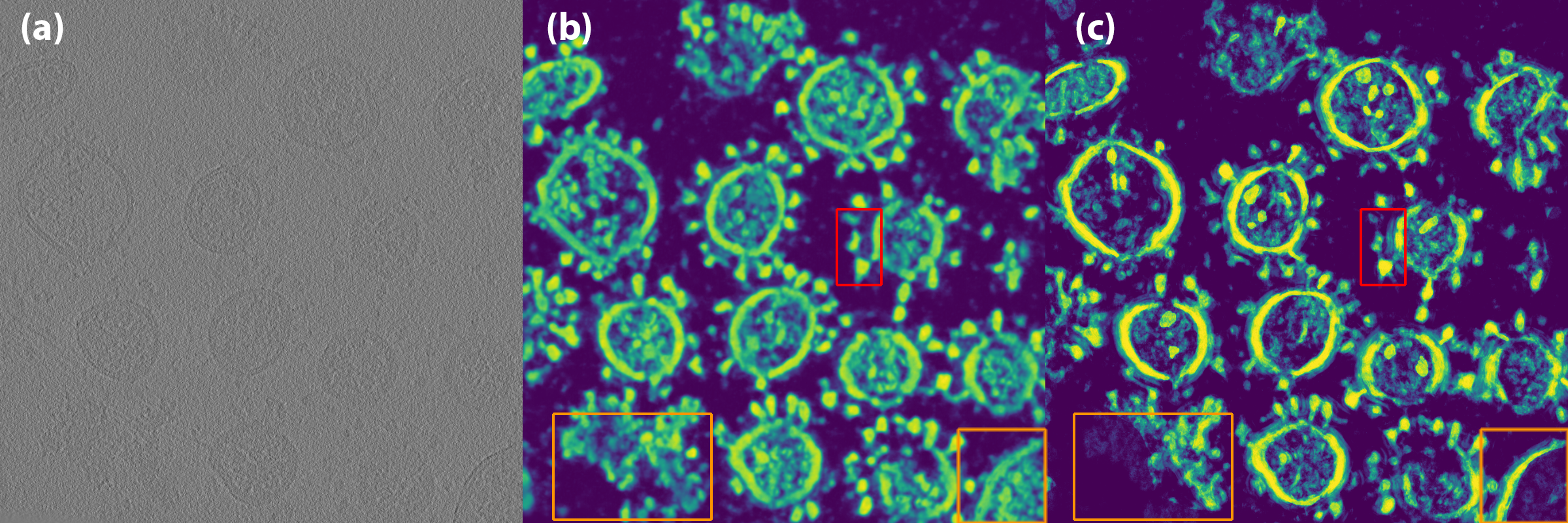}
    \caption{Comparison of original input data (a), automatically segmented data (b), and semi-automatically segmented data (c). In orange boxes are parts of viruses that were not selected by the semi-automatic segmentation but were selected with our approach. In the red boxes are the areas between membrane and spikes which are cleaner in the automatic segmentation.}
    \label{fig:expert1}
\end{figure}

He also confirms that the automatic foreground-background segmentation (AS), i.e. our neural net's predictions, is good. Moreover, he confirms that in some cases, it is even better than the SAS, exposing several structures that were previously not identified. Such structures are presented in \autoref{fig:expert1}. He suggests that some \emph{dust removal} approach could further improve the segmentation by removing particles smaller than the given diameter.

\begin{figure}[h]
    \centering
    \includegraphics[width=0.7\linewidth]{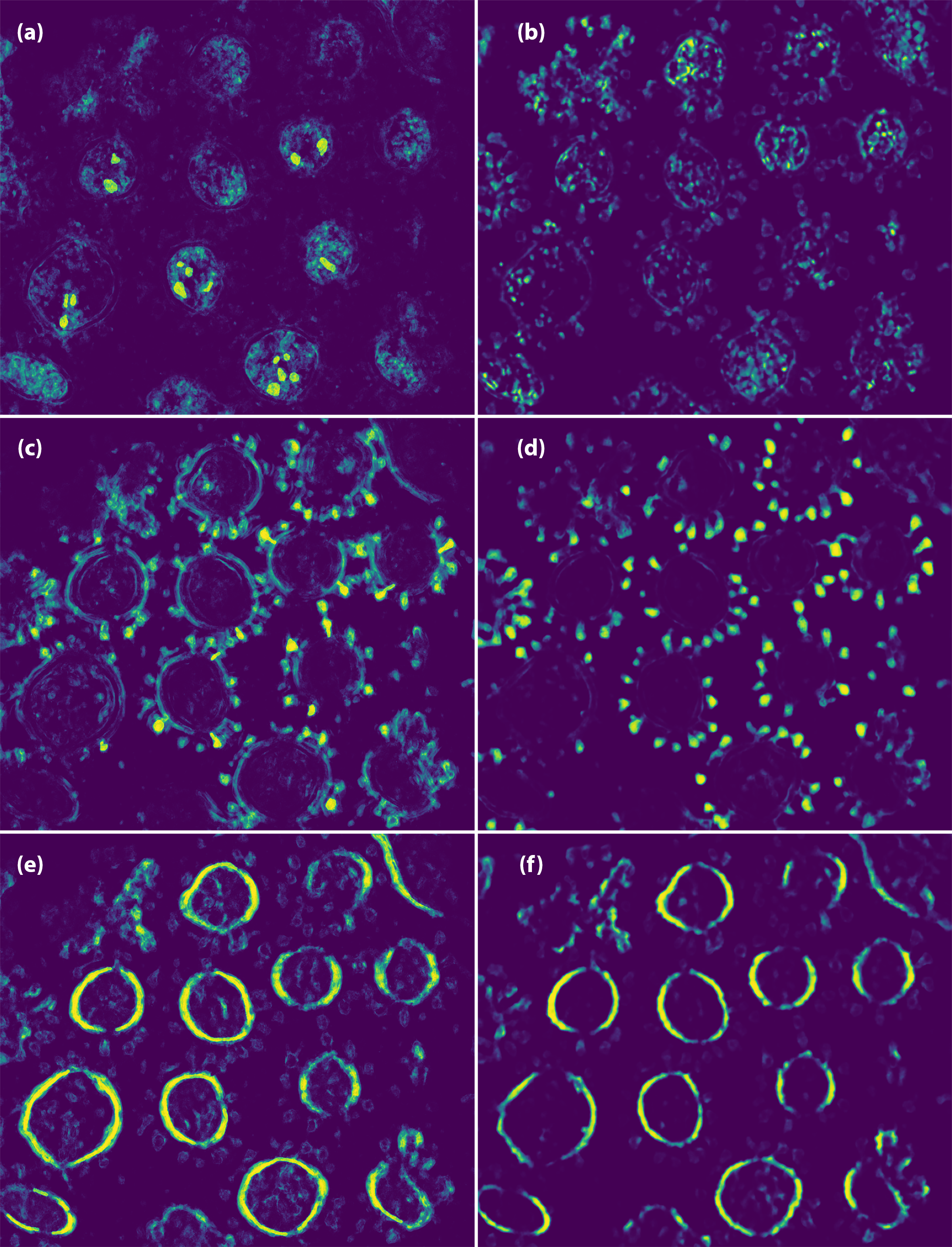}
    \caption{Comparison between four-class segmentation of membranes (bottom) and spikes (middle) and lumen (top) for semi-automatic segmentation (left) and automatic segmentation (right). It is clearly visible that automatic segmentation produces clearer output.}
    \label{fig:expert2}
\end{figure}

After reviewing the four-class semi-automatic segmentations he concluded, that segmentation was performed well but could be further improved. This is true for the lumen structures and the portions of the membrane and spike annotations (see \autoref{fig:expert2} (a), (c), and (e)). It was apparent that there were still some outlines from spikes in some parts of the membrane and vice versa. He suggests addressing this in the future by trying to impose some local limitations to the semi-automatic annotation process. After seeing the automatic segmentation results, he was positively surprised at how well the membrane and spikes segmentation performs (see \autoref{fig:expert2} (d) and (f)). He was delighted that spikes were also present in the \emph{missing wedge} parts, where he did not expect such good results (see right-most virions in \autoref{fig:teaser} and \autoref{fig:results-f-b-vs-four-class} (d)). He confirms that lumen parts segmentation could be further improved, possibly with better semi-automatic segmentations.

We introduced him to three visualization approaches: DVR, MIP, and our approach. He confirmed that he is familiar with the basic DVR visualization techniques and that in his work, he mostly uses the visualization pipeline integrated into the Chimera~\cite{Pettersen2004}. He confirms that basic DVR methods are not suitable for the direct rendering of the data and that our approach is excellent. He supports the claim that the details added from the original tomogram data add to the surface structure's comprehension. He points out that existing visualization packages have additional functionality, which is very useful to the researchers---such as \emph{dust removal}---but is orthogonal to the available visualization settings. He supports that manual transfer function adjustment is beneficial for fine-tuning the output. Furthermore, he suggests adding the options of changing the lighting conditions, which would help explore further details of the specimen.

The second expert is a cell biologist specialized in electron microscopy. He has more than 19 years of experience in EM and 15 years with transmission EM.  He is a team lead of the Electron microscopy laboratory at our University and works with the EM/ET data on average 3 times a month. At first, his work only included biological specimens, but in the last 12 years, he also worked with polymer membranes and catalytic nanoparticles. We have not collaborated with this domain expert and we regard his feedback as fully independent.

His first impression of the data was that it is good data. It is well aligned, the missing wedge is apparent and supports the good alignment, and fiducial markers were mostly removed. He also confirmed that the semi-automatic segmentation (SAS) is good, especially for the time spent per volume. Good fully manual annotation could take up to several days per individual volume. Due to the data's resolution, the segmentation could be further improved by separately selecting individual layers of the lipid bilayer membrane.

He also confirms that automatic foreground-background segmentation (AS) is good. To some extent, it gives even better insight into structures than SAS. While the spikes are sharper in the SAS, they sometimes overlap and merge. On the other hand, the space between spikes and membrane gets smudged often with SAS, while it is much \emph{cleaner} with AS (see red boxes in \autoref{fig:expert1}). Visually, the AS is closer to the real-data due to less saturated areas. He also points out that in many occurrences, the AS finds more structures than the SAS (see orange boxes in \autoref{fig:expert1}, and the value bleeding is more prominent in SAS than in AS, which shows cleaner structures.

We showed him three methods of 3D visualization of the data: DVR, MIP and our approach. He pointed out that he is familiar with even more visualization techniques available in the EM-related software (IMOD~\cite{Mastronarde2017}, Amira Avizo\footnote{\url{https://www.fei.com/software/avizo3d/\%C2\%A0}} and Chimera~\cite{Pettersen2004}). He points out that common volume rendering techniques are not suitable for visualizing such data. He mostly uses transparency-based methods that give better insight into the underlying data structure. Our visualization approach is very appealing to him. It presents structures as solids, while still maintaining the surfaces' textural details revealing the specimen's detailed structure.

Finally, he points out that automation in the segmentation and visualization of EM data is a crucial aspect for accelerating the field's development. Our work addresses this cryo-ET bottleneck and enables scientists from these niches to drastically speed up their research. We present one quote from his statements during the interview: ''\emph{I doubt that you are fully aware about how impactful your technology will be for our field, once it becomes available as a tool}''. It allows quick inspection of the acquired data, making it easier to prioritize which data should be further analyzed sooner. On the other hand, vast datasets were acquired some time ago that were never selected for analysis. With the use of the presented approach, one could revisit these datasets for possible interesting specimens.

The overall feedback from experts was very positive. They confirmed that meaningful and expressive visualization is crucial for understanding the data. They gave us some pointers for possible future extensions of the presented work in the segmentation and the visualization aspects. We agreed that having such a tool in their processing pipeline would definitely speed-up and simplify their analysis process.


\section{Conclusion}
With this work, we have shown how, with a correct approach and a good set of techniques, one can "excavate" and visualize the information present within very tough data with extremely low signal-to-noise ratio, such as cryo-ET. We demonstrate how we can harness the power of deep neural networks to infuse the visualization pipeline with detailed automatic segmentation, yielding high-quality visualization results, where state-of-the-art general volume rendering approaches fail.

While there are limitations to using such a system for a specific pipeline by feeding it with specific segmentations in the training step, it still promises to save days if not weeks of laborious manual segmentation work for obtaining great visual results. By providing even more precise segmentation data---instead of fuzzy semi-automatic segmentations---to our system, the final visualization will be even better. Embedding our system into the cryo-ET pipeline would give domain scientists access to high-quality data visualization at almost no additional effort. The segmentations that domain experts are preparing daily can be used as training input to our system. Moreover, using the data from different laboratories working on similar problems can lead to the preparation of specialized visualization models for specific use-cases, which could be shared back to the research community, benefiting everyone.

The presented system could be further extended to become an end-to-end deep learning system. Not only fuzzy segmentation masks, but also other visualization parameters (e.g., TF parameters and rendering parameters) could be trained for a specific domain. This would include a differentiable volumetric rendering system to allow such endeavor.

Our next step will be to integrate the volume visualization into the data preparation pipelines for the subtomogram averaging process. The results of this signal-to-noise ratio amplification methodology can be fed back into our volume visualization pipeline to further improve the detail an possibly fight the missing wedge artifact.

\bibliographystyle{unsrt}
\bibliography{bibliography}

\end{document}